\documentclass[conference]{IEEEtran}
\IEEEoverridecommandlockouts
\usepackage{amssymb}
\usepackage{amsmath,amssymb,amsfonts}
\usepackage{graphicx}
\usepackage{amsmath}  
\usepackage{amssymb}  
\usepackage{geometry}
\usepackage{mathtools}
\usepackage{newtxmath}
\usepackage{amsmath}
\usepackage{algorithmic}
\usepackage{algorithm}
\usepackage{graphicx}
\usepackage{textcomp}
\usepackage{xcolor}
\usepackage{setspace}

\usepackage[short,c2,nocomma]{optidef}

\usepackage{acronym}
\usepackage{textcomp}

\usepackage{siunitx}
\newgeometry{left =0.673in,right=0.673in, top= 0.75in,bottom=1.1in}

\ifCLASSINFOpdf
\else
\fi

\begin{document}
%
\title{RSMA-Assited and Transceiver-Coordinated 
ICI Management for MIMO-OFDM System}
\author{
    \IEEEauthorblockN{Hengyu Zhang\textsuperscript{1}, Xuehan Wang\textsuperscript{1}, Xu Shi\textsuperscript{1}, Jintao Wang\textsuperscript{1}, Zhaohui Yang\textsuperscript{2}}
    \IEEEauthorblockA{\textsuperscript{1}Beijing National Research Center for Information Science and Technolog (BNRist), Tsinghua University}
    \IEEEauthorblockA{\textsuperscript{2}College of Information Science and Electronic Engineering, Zhejiang University, Hangzhou, China}
    \IEEEauthorblockA{\{{zhanghen23}, {wang-xh21}\}@mails.tsinghua.edu.cn, \{{shi-x},{wangjintao}\}@tsinghua.edu.cn, yang\_zhaohui@zju.edu.cn} 
}
\maketitle

\begin{abstract} 
High-mobility scenarios are becoming increasingly critical in next-generation communication systems. While multiple-input multiple-output orthogonal frequency division multiplexing (MIMO-OFDM) stands as a prominent technology, its performance in such scenarios is fundamentally limited by Doppler-induced inter-carrier interference (ICI). Rate splitting multiple access (RSMA), recognized as a key multiple access technique for future communications, demonstrates superior interference management capabilities that we leverage to address this challenge. In specific, we propose a novel RSMA-assisted and transceiver-coordinated transmission scheme for ICI management in MIMO-OFDM system: (1) At the receiver side, we develop a hybrid successive interference cancellation (SIC) architecture with dynamic subcarrier clustering, which enables parallel intra-cluster and serial inter-cluster processing to balance complexity and performance. (2) At the transmitter~side, we design a matched hybrid precoding through formulated sum-rate maximization, solved via our proposed augmented boundary-compressed particle swarm optimization (ABC-PSO) algorithm for analog phase optimization and weighted minimum mean-square error (WMMSE)-based digital precoding iteration. Simulation results show that our scheme brings effective ICI suppression and enhanced system capacity with controlled complexity.
\end{abstract}


%
\IEEEpeerreviewmaketitle

\section{Introduction}
As a cornerstone of modern wireless communication systems, multiple-input multiple-output orthogonal frequency division multiplexing (MIMO-OFDM) technique provides enhanced spectrum efficiency and dynamic interference management \cite{mimo-ofdm}. However, as future communication systems evolve to support increasingly high-mobility scenarios, the severe Doppler effect will inevitably destroy the orthogonality of OFDM subcarriers and induce inter-carrier interference (ICI) \cite{channel1}. The achievable capacity and transmission reliability of MIMO-OFDM thereby experience significant degradation.  

Recently, rate-splitting multiple access (RSMA) has been proposed to approach the fundamental capacity limit for interference channels, thereby emerging as a viable strategy to confront the ICI caused by the Doppler effect \cite{ofdm-rsma1}. Specifically, RSMA splits the messages for each user into common and private parts, which can flexibly adjust the transmission strategy based on the channel situation, thereby providing more freedom to cope with interference. It also establishes a unified transmission framework that subsumes both non-orthogonal multiple access (NOMA) and space division multiple access (SDMA) paradigms, achieving superior performance in spectral and energy efficiency \cite{rsmasurvey2,yzh}.  

Nevertheless, there exist few researches focusing on MIMO-OFDM system design assisted by RSMA with ICI in doubly dispersive channels. The authors in \cite{mc-rsma1} and \cite{mc-rsma2} conduct the resource allocation for RSMA and achieve excellent performance in MIMO-OFDM system, but they overlook the impact of ICI. Though the authors in \cite{ofdm-rsma1} first exploit RSMA to combat ICI in high-speed mobility scenarios, the proposed solution is only confined to power allocation strategies for subcarriers at the transmitter side, neglecting receiver-side structural design specially for ICI mitigation. This absence of dedicated receiver design limits the achievable performance enhancement.

To address the above limitations, we investigate the interference management strategy for RSMA-assisted MIMO-OFDM systems and propose a novel transceiver-coordinated interference-immune transmission scheme. At the receiver side, conventional RSMA architectures perform successive interference cancellation (SIC) solely at the common-private message level, while insufficient attention is paid to message crosstalk among subcarriers. Thus, we propose a complexity-adaptive hybrid SIC receiver architecture that can effectively suppress ICI and dynamically adjust the overhead. Specifically, subcarriers are partitioned into multiple clusters with intra-cluster parallel SIC and inter-cluster serial SIC. The number of clusters can be modified for the trade-off between performance and complexity. Furthermore, we develop a matching hybrid precoding strategy at the transmitter to jointly cooperate with the receiver structure for interference management. Firstly, the augmented boundary-compressed particle swarm optimization (ABC-PSO) is designed to optimize the analog precoder. The search space of particles is constrained by constant modulus, subsequent feasibility, and quality of service. After that, we propose a weighted minimum mean-square error (WMMSE)-based algorithm for the digital precoder and common rate allocation. Our scheme achieves enhanced system efficiency and manageable~complexity.

\section{System Model}
We consider a MIMO-OFDM downlink system consisting of $K$ single-antenna user equipments (UEs) and a base station (BS) employing a uniform linear array (ULA) with $N_t$ antennas, where the antenna spacing is half of the wavelength. Moreover, there are $N_c$ subcarriers with spacing $\Delta f$ in the OFDM system for resource allocation.
\subsection{Wideband Doubly-Dispersive Channel}
We denote the data vector transmitted at the $m$-th antenna in the frequency domain by $\mathbf{s}_m \in~ \mathbb{C}^{N_c\times 1}$. Time-domain samples can then be generated by inverse fast Fourier transform (IFFT) and adding the cyclic prefix (CP) with length $L$ to combat the multi-path effect as $\tilde{\mathbf{s}}_m\in\mathbb{C}^{(N_c+L)\times 1}$:
\begin{equation}
    \tilde{\mathbf{s}}_m = \mathbf{A}\mathbf{F}^H\mathbf{s}_m, 
    \label{eq:OFDMsignal}
\end{equation}
where $\mathbf{A}\in\mathbb{C}^{(N_c+L)\times N_c}$ and $\mathbf{F}\in\mathbb{C}^{N_c\times N_c}$ represent the CP-addition matrix and the normalized Fourier transform matrix, respectively. Accounting for doubly-dispersive channels, the received signal $\mathbf{r}_k\in\mathbb{C}^{(N_c+L)\times1}$ at the $k$-th UE side is given as 
\begin{equation}
\mathbf{r}_k = \sum_{m=1}^{N_t}\tilde{\mathbf{H}}_{k,m}\tilde{\mathbf{s}}_m + \tilde{\mathbf{z}},
\end{equation}
where $\tilde{\mathbf{z}}\sim\mathcal{CN}(0,\sigma^2\mathbf{I}_{N_c+L})$ denotes the additive white Gaussian noise (AWGN) and $\tilde{\mathbf{H}}_{k,m}\in\mathbb{C}^{(N_c+L)\times(N_c+L)}$ is the channel matrix between the $m$-th antenna of BS and the $k$-th user. Following the doubly-dispersive channel model as \cite{twc1,channel3}, the channel matrix $\tilde{\mathbf{H}}_{k,m}$ can be given as 
\begin{equation}
\tilde{\mathbf{H}}_{k,m} = \sum_{i=1}^{N_p}h_{k,i}e^{j\pi(m-1)\sin\theta_{k,i}}\mathbf{\Delta}^{\nu_{k,i}} \mathbf{\Gamma}_{\tau_{k,i}}. 
\end{equation}

Therein,  $h_{k,i}$, $\tau_{k,i}$, $\nu_{k,i}$, and $\theta_{k,i}$ denote the path gain, time delay, Doppler shift, and angle of departure to the $i$-th path for the $k$-th user, respectively. In addition, $\mathbf{\Delta}\in \mathbb{C}^{(N_c+L)\times (N_c+L)}$ is the Doppler shifting matrix
\begin{equation}\label{eq4}
{\mathbf{\Delta}=\mathrm{diag} [1,e^{j2\pi/F_s},e^{j4\pi /F_s},...,e^{j2\pi(N_c+L-1)/F_s}]},
\end{equation}
where $F_s=N_c\triangle f$ denotes the sampling frequency. The time delay matrix $\mathbf{\Gamma}_{\tau_{k,i}}\in \mathbb{C}^{(N_c+L)\times (N_c+L)}$ can be given by
\begin{equation}
\left [\mathbf{\Gamma}_{\tau_{k,i}}\right ]_{p,q} = g(\frac{p-q}{F_s}-\tau_{k,i}), \ p,q\in\{1,2,...,N_c+L \},   
\end{equation}
where $g(\cdot)$ is the practical pulse shaping at the transmitter side \cite{oddm}. Then, the frequency domain symbols can be derived by $\mathbf{y}_k\in\mathbb{C}^{N_c\times 1}$ after the corresponding operation:

\begin{equation}
\begin{aligned}
\mathbf{y}_k &= \mathbf{F}\mathbf{B}(\sum_{m=1}^{N_t}\tilde{\mathbf{H}}_{k,m}\tilde{\mathbf{s}}_m + \tilde{\mathbf{z}})=\sum_{m=1}^{N_t}\mathbf{H}_{k,m}\mathbf{s}_m+\mathbf{z},
\end{aligned}
\end{equation}
where $\mathbf{B}\in\mathbb{C}^{N_c\times(N_c+L)}$ denotes the CP-removal matrix and $\mathbf{z}=\mathbf{F}\mathbf{B}\tilde{\mathbf{z}}$. To simplify the expression, $\mathbf{H}_{k,m}\triangleq\mathbf{F}\mathbf{B}\tilde{\mathbf{H}}_{k,m}\mathbf{A}\mathbf{F}^H$ represents the equivalent channel matrix in the frequency domain between the $m$-th antenna of BS and the $k$-th user.
\subsection{RSMA Transmission with ICI}
Let $d_{c,n}$ and $d_{k,n}$ denote the common message and the $k$-th UE's private message at the $n$-th subcarrier, respectively. Assuming $K$ radio frequency (RF) chains, the transmitted symbols $\mathbf{x}_n \in \mathbb{C}^{N_t\times 1}$ at the $n$-th subcarrier can be written as
\begin{equation}
\begin{aligned}
     \mathbf{x}_{n} &= \mathbf{F}_{\text{RF}}(\mathbf{F}_{\text{BB},c,n}d_{c,n}+\sum_{k=1}^{K}\mathbf{F}_{\text{BB},k,n}d_{k,n})\\
     &= \underbrace{\mathbf{F}_{\text{RF}}\mathbf{F}_{\text{BB},c,n}d_{c,n}}_{\mathbf{x}_{c,n}}+\underbrace{\mathbf{F}_{\text{RF}}\sum_{k=1}^{K}\mathbf{F}_{\text{BB},k,n}d_{k,n}}_{\mathbf{x}_{p,n}},
\end{aligned}
\end{equation}
where $\mathbf{F}_{\text{RF}}\in \mathbb{C}^{N_t\times K},\mathbf{F}_{\text{BB},c,n},\mathbf{F}_{\text{BB},k,n}\in \mathbb{C}^{K\times 1}$ represents the analog precoder, digital precoding vectors for the common and private data stream at the $n$-th subcarrier, respectively. Thus, we can obtain $\mathbf{y}_k$ by substituting $\mathbf{s}_m=\left [\left[\mathbf{x}_{1}\right]_m,\left[\mathbf{x}_{2}\right]_m...\left[\mathbf{x}_{N_c}\right]_m \right ]^T$ to (6).

Based on the downlink RSMA transmission principle, each UE needs to decode the common message first. We can rewrite the ideal received message $y_{c,k,n}$ at the $n$-th subcarrier of the $k$-th user according to (6) and (7).
\begin{equation}
\small
\begin{aligned}
y_{c,k,n}&=\mathbf{h}_{k,n,n}^H\mathbf{x}_{c,n}+\sum_{n'\ne n} \mathbf{h}_{k,n,n'}^H\mathbf{x}_{c,n'}+\sum_{n'=1}^{N_c} \mathbf{h}_{k,n,n'}^H \mathbf{x}_{p,n'}\\
&=w_{k,n,n}d_{c,n}+\sum_{n'\ne n}w_{k,n,n'}d_{c,n'}+\sum_{n'=1}^{N_c}\sum_{k'=1}^{K}v_{k,k',n,n'}d_{k',n'},
\end{aligned}
\end{equation}
where $\mathbf{h}_{k,n,n'} = \left [ \left[\mathbf{H}_{k,1}\right]_{n,n'},\left[\mathbf{H}_{k,2}\right]_{n,n'},...,\left[\mathbf{H}_{k,N_t}\right]_{n,n'}\right ] ^H$ is the channel vector reflecting on the effect from the $n'$-th subcarrier to the $n$-th one. The noise term has been omitted for expression simplification. Besides, equivalent transmission coefficients $w_{k,n,n'}$ and $v_{k,k',n,n'}$ can be respectively given as
\begin{equation}
w_{k,n,n'}= \mathbf{h}_{k,n,n'}^H \mathbf{F}_{\text{RF}}\mathbf{F}_{\text{BB},c,n'},
\end{equation}
\begin{equation}
v_{k,k',n,n'}= \mathbf{h}_{k,n,n'}^H \mathbf{F}_{\text{RF}}\mathbf{F}_{\text{BB},k',n'}.
\end{equation}

\begin{figure*}
\centering
\includegraphics[width=0.677\linewidth]{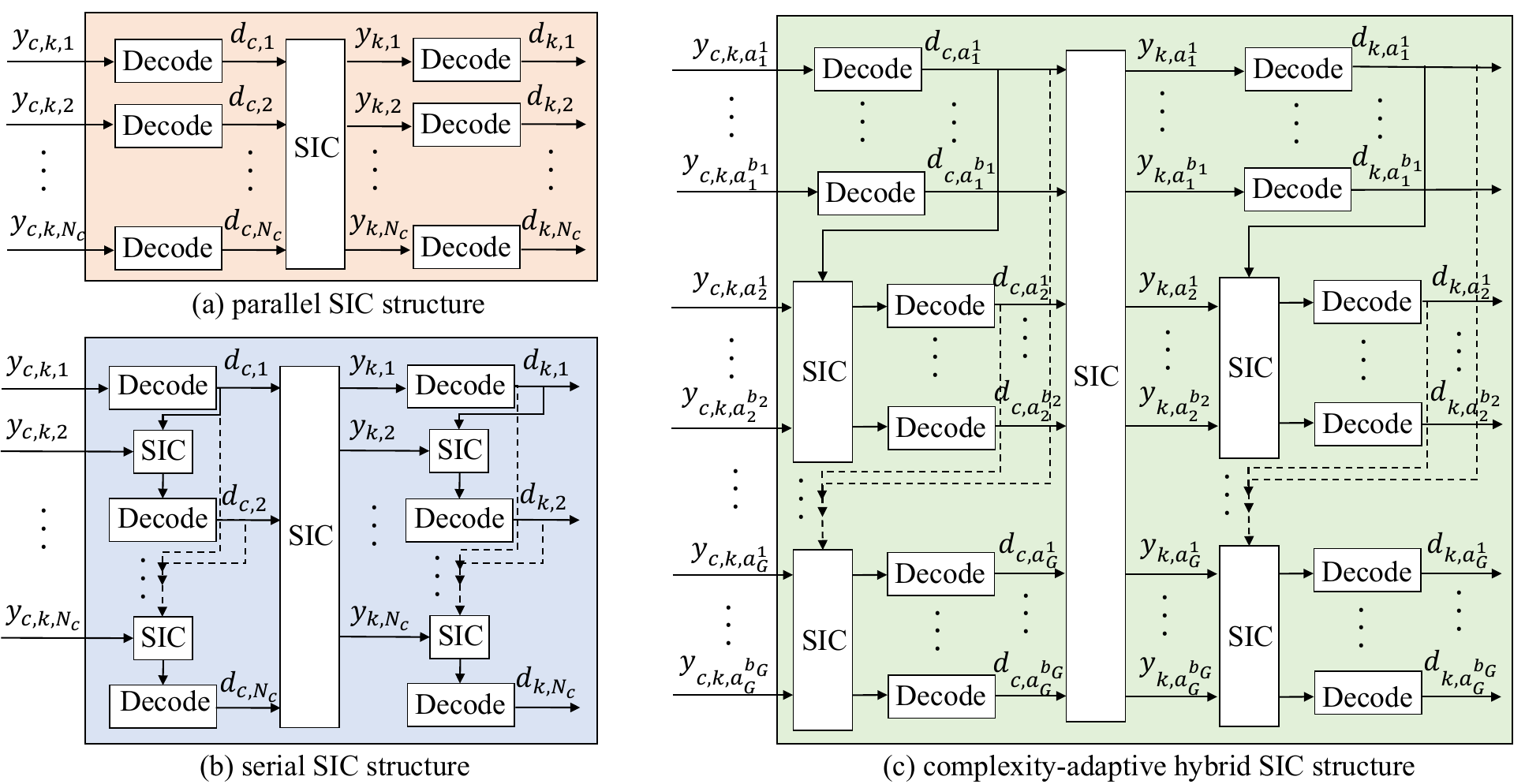}
\caption{(a) Parallel SIC, overlooking ICI and thus with unsatisfactory performance. (b) Serial SIC, suffering from huge computational complexity. (c) Our proposed hybrid SIC structure, with trade-off between complexity and performance adaptively.}
\label{fig_3}    
\end{figure*}
Evidently, the interference of $y_{c,k,n}$ comes from the common messages of other subcarriers and all private messages. Assuming the perfect SIC of common messages as \cite{ofdm-rsma1,rsmasurvey2}, the received signal $y_{k,n}$ at the $n$-th subcarrier of the $k$-th user can be given by
\begin{equation}
\small
\begin{aligned}
y_{k,n}&=v_{k,k,n,n}d_{k,n}+\sum_{n'\ne n}v_{k,k,n,n'}d_{k,n'}+\sum_{n'=1}^{N_c}\sum_{k'\ne k}v_{k,k',n,n'}d_{k',n'},
\end{aligned}
\end{equation}
where the interference is not only from own messages of  other subcarriers, but also others' messages of all subcarriers.
\section{Complexity-Adaptive Hybrid SIC Receiver}
The severe Doppler effect significantly degrades MIMO-OFDM system performance due to ICI. Conventional RSMA receivers employ the parallel SIC architecture illustrated in Fig. 1(a), where each subcarrier processes its messages independently \cite{ofdm-rsma1}. Interference from common messages is only eliminated through parallel SIC after complete decoding of all common messages. While this structure offers low complexity, its lack of dedicated ICI mitigation mechanisms inevitably compromises system performance. A straightforward solution is to adopt the serial SIC structure depicted in Fig. 1(b), sequentially decoding subcarrier information while utilizing previously decoded messages to partially cancel ICI. However, this approach suffers from prohibitively high complexity, rendering it impractical for rapidly time-varying channels.

To overcome the limitations of architectures above, we propose a novel complexity-adaptive hybrid SIC receiver, as illustrated in Fig. 1(c). This new paradigm transforms the conventional parallel or serial scheme into an integrated hierarchical structure. The core innovation lies in a flexible configuration process that partitions subcarriers into $G$ clusters $\left \{\kappa_1,\kappa_2,...,\kappa_{G} \right \}$. Assuming that the $g$-th cluster $\kappa_g$ contains $b_g$ subcarriers, $a_{g}^{i}$ denotes the index of the $i$-th subcarrier in $\kappa_g$. Thus, we can describe the $g$-th cluster as $\kappa_g=\left \{a_{g}^{1},a_{g}^{2},...,a_{g}^{b_g} \right \}$. The subcarriers within the same cluster undergo parallel SIC processing, while serial SIC is arranged across clusters. This bi-level hierarchical architecture creates a continuous design space spanning from pure parallel ($G=1$) to fully serial ($G=N_c$) implementations, governed by the adaptability parameter $G$ that precisely controls the complexity-performance trade-off. Through appropriate selection of $G$ and cluster configurations, it can leap in interference management flexibility, particularly crucial for high-mobility scenarios with time-varying channels.

Based on the hybrid SIC structure, the average received power when decoding $y_{c,k,n}$ is denoted by $T_{c,k,n}$ for the $k$-th UE and $n$-th subcarrier
\begin{equation}
\small
    T_{c,k,n}  =   \left \vert w_{k,n,n} \right \vert^2 + \underbrace{ \sum_{n'\in\mathfrak{K}_n }^{\ne n} \left \vert w_{k,n,n'} \right \vert^2 + \sum_{n'=1}^{N_c}\sum_{k'=1}^{K} \left \vert v_{k,k',n,n'} \right \vert^2 + \sigma^2}_{I_{c,k,n}}, 
\end{equation}
where $I_{c,k,n}$ denotes the average interference power plus noise power when decoding $y_{c,k,n}$. Assuming that the $n$-th subcarrier belongs to the $\tilde{n}$-th cluster $\kappa_{\tilde{n}}$, $\mathfrak{K}_n=\left \{ \kappa_{\tilde{n}},\kappa_{\tilde{n}+1},...,\kappa_{G}\right \}$ denotes the set of interference clusters to the $n$-th subcarrier. Part of the interference among clusters has been removed owing to our hybrid structure.

Similarly, we can obtain the average received power $T_{k,n}$ when decoding $y_{k,n}$ as follows:
\begin{equation}
\small
    T_{k,n}  =   \left \vert v_{k,k,n,n} \right \vert^2 + \underbrace{ \sum_{n'\in\mathfrak{K}_n }^{\ne n} \left \vert v_{k,k,n,n'} \right \vert^2 + \sum_{n'=1}^{N_c}\sum_{k'\ne k} \left \vert v_{k,k',n,n'} \right \vert^2 + \sigma^2}_{I_{k,n}},
\end{equation}
where $I_{k,n}$ denotes the average interference power plus noise power when decoding $y_{k,n}$.

\section{Matching Transmitter with Hybrid Precoding}
Besides the specialized receiver architecture, a matching transmitter mechanism is essential for effective ICI mitigation. We formulate a system sum-rate maximization problem at the transmitter side based on the principles of hybrid SIC reception. Subsequently, we develop an iterative optimization algorithm to alternatively design analog and digital precoders, finally establishing an efficient hybrid precoding scheme.

\subsection{Problem Formulation}
According to (12) and (13), we can obtain the SINRs of common messages $\gamma_{c,k,n}$ and private messages $\gamma_{k,n}$ as
\begin{equation}
\gamma_{c,k,n}\triangleq\left \vert w_{k,n,n} \right \vert^2 I^{-1}_{c,k,n},\quad \gamma_{k,n} \triangleq \left \vert v_{k,k,n,n} \right \vert^2  I^{-1}_{k,n}.
\end{equation}

As a result, the achievable rates for the common message and private message of the $k$-th UE at the $n$-th subcarrier can be respectively denoted by $R_{c,k,n}$ and $R_{k,n}$
\begin{equation}
R_{c,k,n}= \log_2(1+\gamma_{c,k,n}),\quad R_{k,n} = \log_2(1+\gamma_{k,n}).
\end{equation}

To guarantee that each UE can successfully decode the common message in all subcarriers, the achievable rate for the common message should be limited by $ R_{c,n} = \min_{k} R_{c,k,n}$, and allocated to each UE as $\sum_{k=1}^K C_{k,n} \leq R_{c,n}$. Therein, $C_{k,n}$ is the rate of the $n$-th subcarrier assigned to the $k$-th UE. 

Based on that, the sum-rate maximization problem of OFDM-RSMA system can be written as\footnote{Considering the extremely high freedom of cluster configurations $\kappa_g$, it is almost impossible to obtain the optimal cluster division with limited computation resources. Moreover, it is necessary to inform both transmitter and receiver of the optimal cluster division, which leads to too heavy overhead. Therefore, we do not intend to optimize it and assume a fixed cluster division that both transmitter and receiver know in advance in this paper.}:
\begin{maxi!}<b>
{\mathbf{F}_{\text{RF}},\mathbf{F}_{\text{BB},n},\mathbf{C}}{R_\text{sum}=\sum_{n=1}^{N_c}R_{c,n}+ \sum_{n=1}^{N_c}\sum^K_{k=1}R_{k,n} \label{eq:optProblem1}}
{\label{eq:optProblem}}{}
\addConstraint{R_{c,n}}{\leq R_{c,k,n}, \quad \forall k,n, \label{eq:optProblem2}}
\addConstraint{\sum_{n=1}^{N_c}\left \| \mathbf{F}_{\text{RF}}\mathbf{F}_{\text{BB},n} \right \|_{F}^2  {\leq P_t, \label{eq:optProblem3}}}
\addConstraint{\sum_{n=1}^{N_c} (C_{k,n}+R_{k,n}){\geq R_k^{min},  \quad \forall k,\label{eq:optProblem4}}}
\addConstraint{ \sum_{k=1}^K C_{k,n} \leq R_{c,n}, \quad \forall n,\label{eq:optProblem4}}
\addConstraint{| \mathbf{F}_{\text{RF}}|_{i,j} = 1,\quad \forall  i,j \label{eq:optProblem3}}
\end{maxi!}
where $\mathbf{F}_{\text{BB},n}=\left [ \mathbf{F}_{\text{BB},c,n},\mathbf{F}_{\text{BB},1,n},...,\mathbf{F}_{\text{BB},K,n}\right]\in \mathbb{C}^{K\times (K+1)}$ denotes the digital precoder of the $n$-th subcarrier and $\mathbf{C}=\{C_{k,n}\vert\forall k,n\}$. The objective function (16a) represents the sum rate of the system. Constraint (16b) ensures each UE can decode the common message. (16c) defines the power constraint $P_t$ of the overall system, and (16d) ensures that each UE should achieve the sum rate of $R_k^{min}$ at least. Moreover, (16e) represents the allocation of the common rate. Finally, (16f) ensures the constant modulus of the analog precoder.

Evidently, the closed-form optimal solution to (16) is almost impossible to directly obtain since it is non-convex. Thus, we intend to iteratively optimize $\mathbf{F}_{\text{RF}}$ and $\mathbf{F}_{\text{BB},n}$, gradually approaching the suboptimal solution of hybrid precoding.

\subsection{ABC-PSO Algorithm for Analog Precoding}
Given fixed $\mathbf{F}_{\text{BB},n}$, it is still hard to derive the closed-form solution of $\mathbf{F}_{\text{RF}}$ due to its high-dimensional search space~\cite{BC-PSO}. Inspired by particle swarm optimization (PSO) \cite{PSO}, we simulate $I$ random particles to search the optimal $\mathbf{F}_{\text{RF}}$. Let $\mathbf{A}_l\in \mathbb{C}^{N_t \times K}$ and $\mathbf{V}_l\in \mathbb{C}^{N_t \times K}$ denote the position (i.e., currently searched $\mathbf{F}_{\text{RF}}$) and velocity of the $l$-th particle, respectively. Each particle updates its position and velocity in each iteration following
\begin{equation}
\begin{aligned} 
[\mathbf{V}_l]_{i,j}=&\ \omega[\mathbf{V}_l]_{i,j}+c_1\beta_1([\mathbf{P}_{\text{best},l}]_{i,j}-[\mathbf{A}_l]_{i,j})\\& \ +c_2\beta_2([\mathbf{G}_{\text{best}}]_{i,j}-[\mathbf{A}_l]_{i,j}) ,\\  
[\mathbf{A}_l]_{i,j} = &\ [\mathbf{A}_l]_{i,j} +[\mathbf{V}_l]_{i,j},\ (i=1,..,N_t,\ j=1,...,K),
\end{aligned} 
\end{equation}
where $\omega$, $c_1$, and $c_2$ denote the convergence ratio, cognitive ratio, and social ratio, respectively. Besides, $\beta_1$ and $\beta_2$ represent two constants randomly generated between 0 and 1. $\mathbf{P}_{\text{best},l}$ is the best position of the $l$-th particle during its iteration history, while $\mathbf{G}_{\text{best}}$ is so far the globally best position. 

Meanwhile, the update of $\mathbf{A}_l$ needs to satisfy the power constraint (16c) and guarantee that there exists a feasible common rate allocation $\mathbf{C}$ to satisfy (16d). Indeed, we can exploit the water-filling to successfully allocate the common rate only if the following inequality holds:
\begin{equation}
    \sum_{k=1}^{K}\max\{R_k^{min}-\sum_{n=1}^{N_c}R_{k,n},0\}\leq \sum_{n=1}^{N_c}R_{c,n}.
\end{equation}

These constraints make it difficult to directly apply PSO algorithm. Thus, we design the ABC-PSO to bound the search space of particles considering these constraints. Specifically, an augmented fitness function $\mathcal{F}(\mathbf{A}_l)$ is introduced to evaluate the position of the $l$-th particle:
\begin{equation}
\label{pilot}
\mathcal{F}(\mathbf{A}_l)=\left\{\begin{aligned} 
&0,\quad \text{if (16c) or (18) is not satisfied},\\  
&R_\text{sum}=\sum_{n=1}^{N_c}R_{c,n}+ \sum_{n=1}^{N_c}\sum^K_{k=1}R_{k,n}, \ \text{otherwise}.
\end{aligned}\right. 
\end{equation}
\begin{algorithm}[t]
\setstretch{0.8}
    \caption{ABC-PSO Algorithm}
    \label{alg:AOA}
    \begin{algorithmic}[1]
        \STATE \textbf{Input:} $K$, $N_t$, $N_c$, $P_t$, $R_k^{\text{min}}$, $\mathbf{H}_{k,m}$,
        \\ \quad  \quad \ \ \ \{$I$, $T_{\text{max}}$, $c_1$, $c_2$, $\omega_{\text{max}}$, $\omega_{\text{min}}$\}.
        \STATE \textbf{Output:} $\mathbf{F}_{\text{RF}}$.
            \STATE \textbf{Initialize:} $\mathbf{V}_l$ and $\mathbf{A}_l$.
            \STATE Calculate $\mathcal{F}(\mathbf{A}_l)$ based on (19), then find $\{\mathbf{P}_{\text{best},l},\mathbf{G}_{\text{best}}\}$.
            \FOR {$t$ = 1, 2, ... , $T_{\text{max}}$}
                \STATE $\omega = \omega_{\text{max}}-\frac{t}{T_{\text{max}}}(\omega_{\text{max}}-\omega_{\text{min}})$, $d_{\text{out}}=1, d_{\text{in}}=\frac{t}{T_{\text{max}}}$
                \FOR {$l$ = 1, 2, ... , $I$}
                    \FOR {$i$ = 1, 2, ... , $N_t$}
                        \FOR {$j$ = 1, 2, ... , $K$}
                        \STATE Update $[\mathbf{V}_l]_{i,j}$ and $[\mathbf{A}_l]_{i,j}$ as (17).
                        \IF{$\vert[\mathbf{A}_l]_{i,j} \vert<d_\text{in}$}
                        \STATE $[\mathbf{A}_l]_{i,j}=d_\text{in}\frac{[\mathbf{A}_l]_{i,j}}{\vert[\mathbf{A}_l]_{i,j}}$
                        \ENDIF
                        \IF{$\vert[\mathbf{A}_l]_{i,j} \vert>d_\text{out}$}
                        \STATE $[\mathbf{A}_l]_{i,j}=d_\text{out}\frac{[\mathbf{A}_l]_{i,j}}{\vert[\mathbf{A}_l]_{i,j}}$
                        \ENDIF
                        \IF{$\vert[\mathbf{P}_{\text{best},l}]_{i,j} \vert<d_\text{in}$}
                        \STATE $[\mathbf{P}_{\text{best},l}]_{i,j}=d_\text{in}\frac{[\mathbf{P}_{\text{best},l}]_{i,j}}{\vert[\mathbf{P}_{\text{best},l}]_{i,j}}$
                        \ENDIF
                        \ENDFOR
                    \ENDFOR 
                \ENDFOR
            \STATE Obtain $\mathcal{F}(\mathbf{A}_l)$ based on (19) and update $\{\mathbf{P}_{\text{best},l},\mathbf{G}_{\text{best}}\}$.    
            \ENDFOR
            \STATE $\mathbf{F}_{\text{RF}}=\mathbf{G}_{\text{best}}$ 
    \end{algorithmic}
\end{algorithm}

That is to say, if the position of the $l$-th particle leads to an unfeasible solution, we set its fitness function as zero. Besides, $\mathbf{F}_{\text{RF}}$ also needs to satisfy the constant modulus (CM) constraint as (16f). We decide to gradually narrow the search space of the particle along with iteration, finally meeting the CM requirement. Specifically, during the $t$-th iteration of the particle search, $\vert[\mathbf{A}_l]_{i,j} \vert$ should be limited in the range of $\left[\frac{t}{T_\text{max}},1\right]$, where $T_\text{max}$ is the maximum number of iterations. The position $\mathbf{A}_l$ will be adjusted within the search space if beyond the boundary. The details of the ABC-PSO algorithm are presented in \textbf{Algorithm 1}.

\subsection{WMMSE-based Algorithm for Digital Precoding}
After optimizing a fixed $\mathbf{F}_{\text{RF}}$ through ABC-PSO, the constraint of constant modulus (16f) can be eliminated temporarily. In this way, we design a WMMSE-based algorithm to transform the coupled sum-rate expression into a concise linear expression, then obtaining the solution of digital precoding.    

Let $g_{c,k,n}$ and $g_{k,n}$ be the $k$-th UE's scalar equalizers for common and private messages at the $n$-th subcarrier, respectively. Thus, the corresponding estimates of $d_{c,n}$ and $d_{k,n}$ equal to $\hat{d}_{c,k,n}=g_{c,k,n}y_{c,k,n}$ and $\hat{d}_{k,n}=g_{k,n}y_{k,n}$. The common and private mean-square errors (MSEs) at the $n$-th subcarrier are defined as $\varepsilon_{c,k,n} \triangleq \mathbb{E}\left\{ \left \vert \hat{d}_{c,k,n} - d_{c,n} \right \vert^2 \right\}$
and $\varepsilon_{k,n} \triangleq \mathbb{E}\left\{ \left \vert \hat{d}_{k,n} - d_{k,n} \right \vert^2 \right\}$ respectively:
\begin{equation}
    \varepsilon_{c,k,n} = \left \vert g_{c,k,n} \right \vert^2 T_{c,k,n} - 2\mathfrak{R}\left\{ g_{c,k,n} w_{k,n,n} \right\} + 1,    
\end{equation}
\begin{equation}
    \varepsilon_{k,n} = \left \vert g_{k,n} \right \vert^2 T_{k,n} - 2\mathfrak{R}\left\{ g_{k,n} v_{k,k,n,n} \right\} + 1.    
\end{equation}

We can obtain optimum minimum MSE (MMSE) equalizers by calculating $\frac{\partial \varepsilon_{c,k,n}}{\partial g_{c,k,n}} = 0$ and $\frac{\partial \varepsilon_{k,n}}{\partial g_{k,n}} = 0$ as
\begin{equation}
g_{c,k,n}^{\text{MMSE}} = T_{c,k,n}^{-1} w_{k,n,n}^H,\quad g_{k,n}^{\text{MMSE}} = T_{k,n}^{-1} v_{k,k,n,n}^H.
\end{equation}

Substituting (22) into (20) and (21), MMSEs can be given as
\begin{equation}
\varepsilon^{\text{MMSE}}_{c,k,n} = T_{c,k,n}^{-1} I_{c,k,n}, \quad \varepsilon_{k,n}^{\text{MMSE}} = T_{k,n}^{-1} I_{k,n}.
\end{equation}

As a result, we can rewrite the SINRs as $\gamma_{c,k,n}= \frac{1-\varepsilon^{\text{MMSE}}_{c,k,n}}{\varepsilon^{\text{MMSE}}_{c,k,n}}$ and $\gamma_{k,n} = \frac{1-\varepsilon_{k,n}^{\text{MMSE}}}{\varepsilon_{k,n}^{\text{MMSE}}}$. Moreover, the achievable rates can also be updated as $R_{c,k,n}=-\log_2(\varepsilon^{\text{MMSE}}_{c,k,n})$ and $R_{k,n} = -\log_2(\varepsilon_{k,n}^{\text{MMSE}})$. Then we introduce 
$u_{c,k,n},u_{k,n} > 0$ as weights for the $k$-th UE's MSEs at the $n$-th subcarrier and adopt the augmented weighted MSEs (AWMSEs) as
\begin{equation}
    \zeta_{c,k,n} = u_{c,k,n} \varepsilon_{c,k,n} - \log_2\left(u_{c,k,n}\right),   
\end{equation}
\begin{equation}
    \zeta_{k,n} = u_{k,n} \varepsilon_{k,n} - \log_2\left(u_{k,n}\right).
\end{equation}

To obtain the smallest AWMSEs, we first solve $\frac{\partial \zeta_{c,k,n}}{\partial g_{c,k,n}} = 0$ and $\frac{\partial \zeta_{k,n}}{\partial g_{k,n}} = 0$, getting and substituting the optimum equalizers $g_{c,k,n}^{*}=g_{c,k,n}^{\text{MMSE}}$ and $g_{k,n}^{*}=g_{k,n}^{\text{MMSE}}$ as 
\begin{equation}
\zeta_{c,k,n}\left(g_{c,k,n}^{\text{MMSE}}\right) = u_{c,k,n}\varepsilon^{\text{MMSE}}_{c,k,n} - \log_2 \left(u_{c,k,n}\right),  
\end{equation}
\begin{equation}
\zeta_{k,n}\left(g_{k,n}^{\text{MMSE}}\right) = u_{k,n} \varepsilon^{\text{MMSE}}_{k,n} - \log_2\left(u_{k,n} \right).  
\end{equation}

After that, $u_{c,k,n}^{*}=u_{c,k,n}^{\text{MMSE}}\triangleq \left(\varepsilon^{\text{MMSE}}_{c,k,n}\right)^{-1}/\text{ln2}$ and $u_{k,n}^{*}=u_{k,n}^{\text{MMSE}}\triangleq \left(\varepsilon^{\text{MMSE}}_{k,n}\right)^{-1}/\text{ln2}$ can be derived from $\frac{\partial \zeta_{c,k,n}}{\partial u_{c,k,n}} = 0$ and $\frac{\partial \zeta_{k,n}}{\partial u_{k,n}} = 0$. The optimum AWMSEs can be given as
\begin{equation}
\zeta^{\text{MMSE}}_{c,k,n} =\frac{1}{\text{ln2}}+\log_2\left(\text{ln2}\right)-R_{c,k,n} =\lambda -R_{c,k,n},  
\end{equation}
\begin{equation}
\zeta^{\text{MMSE}}_{k,n} = \frac{1}{\text{ln2}}+\log_2\left(\text{ln2}\right)-R_{k,n}=\lambda -R_{k,n},  
\end{equation}
where $\lambda \triangleq \frac{1}{\text{ln2}}+\log_2\left(\text{ln2}\right)$. Equations (28) and (29) indicate the relationship between the achieved rates and AWMSEs. We can derive the deterministic version of the AWMSE minimization problem as follows:
\begin{mini!}<b>
{\boldsymbol{\zeta}_c,\mathbf{F}_{\text{BB},n},\mathbf{C},\mathbf{U},\mathbf{G}}
{\zeta_\text{sum}=\sum_{n=1}^{N_c}\zeta_{c,n}+ \sum_{n=1}^{N_c}\sum^K_{k=1}\zeta_{k,n}\label{eq:optProblem_31}}
{\label{eq:optProblem_3}}{}
\addConstraint{ \zeta_{c,k,n}}{\leq  \zeta_{c,n},\quad \forall k,n}
\addConstraint{\sum_{n=1}^{N_c}\left \| \mathbf{F}_{\text{RF}}\mathbf{F}_{\text{BB},n} \right \|_{F}^2  {\leq P_t,}}
\addConstraint{  \sum_{n=1}^{N_c} \zeta_{k,n}}{\leq N_c\lambda - R_k^{\text{min}}+\sum_{n=1}^{N_c} C_{k,n},\quad \forall k,}
\addConstraint{\zeta_{c,n}}{\leq \lambda - \sum_{k=1}^K C_{k,n},\quad \forall n,}
\end{mini!}
where $\boldsymbol{\zeta}_c=\{\zeta_{c,n}\vert\forall n\}$ represents the common AWMSE and $\mathbf{U}=\{u_{c,k,n},u_{k,n}\vert\forall k,n\}$,$\mathbf{G}=\{g_{c,k,n},g_{k,n}\vert\forall k,n\}$. Substituting the relation between the achieved rates and AWMSEs to problem (16), we can modify the constraints as (30b)-(30e). Although the AWMSE minimization problem is still non-convex, we can fix $\mathbf{F}_{\text{BB},n}$ to optimize $\{\mathbf{U},\mathbf{G}\}$ or vice versa, thus performing an iterative optimization. Specifically, if $\mathbf{F}_{\text{BB},n}$ is fixed, we can obtain the optimum $\{\mathbf{U},\mathbf{G}\}$ according to $\{u_{c,k,n}^{\text{MMSE}},u_{k,n}^{\text{MMSE}}\}$ and $\{g_{c,k,n}^{\text{MMSE}},g_{k,n}^{\text{MMSE}}\}$. With the updated $\{\mathbf{U},\mathbf{G}\}$, the optimization problem of (30) degrades into a convex Quadratically Constrained Quadratic Program (QCQP) and can be solved by CVX of Matlab\footnote{Indeed, this hybrid precoding algorithm can also be modified for NOMA and SDMA, only with the expression of the objective function and some constraints slightly changed.}.

\section{Simulation Results}
In this section, simulation results are provided to demonstrate the performance of the proposed scheme under the setting of $\{K=2,N_c=16,N_t=16\}$. We generate 500 random channels for simulation based on the doubly-dispersive channel in Section II. The channel parameters mainly follow \cite{channel1,channel3}. In specific, $f_c,\triangle f, N_p$, and $L$ are set 4 GHz, 15 kHz, 50, and 6, respectively. In addition, the path gain $h_{k,i}$ of the $i$-th path for the $k$-th user is assumed ${h_{k,i}}\sim\mathcal{CN}(0,\frac{1}{N_{p}})$. The time delay $\tau_{k,i}$ is randomly generated as $\mathcal{U}(0,\frac{4}{F_s})$, while Doppler shift is uniformly distributed over $\mathcal{U}(-\nu_{max},\nu_{max})$. Therein, we can obtain the maximum Doppler shift as $\nu_{max}=\frac{vf_c}{c}$, where $c$ denotes the speed of light and $v=360$ km/h represents the maximum relative speed. Moreover, we select the raised cosine function with a roll-off factor of 0.4 as $g(t)$. The parameters for ABC-PSO are set following $\{I=400, T_{\text{max}}=100, c_1=c_2=1.4, \omega_{\text{max}}=0.9, \omega_{\text{min}}=0.4\}$.

\begin{figure}[!t]
\centering
\includegraphics[width=0.62\linewidth]{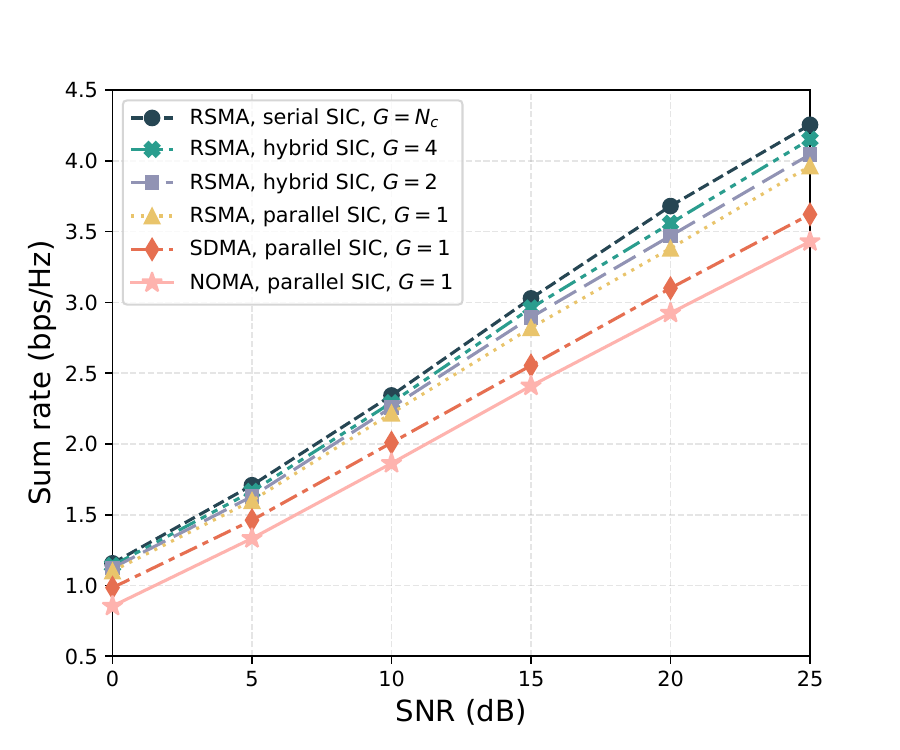}
\caption{Performance of different multiple access techniques versus SNR.}
\label{fig_3}
\end{figure}

\begin{figure}[!t]
\centering
\includegraphics[width=0.62\linewidth]{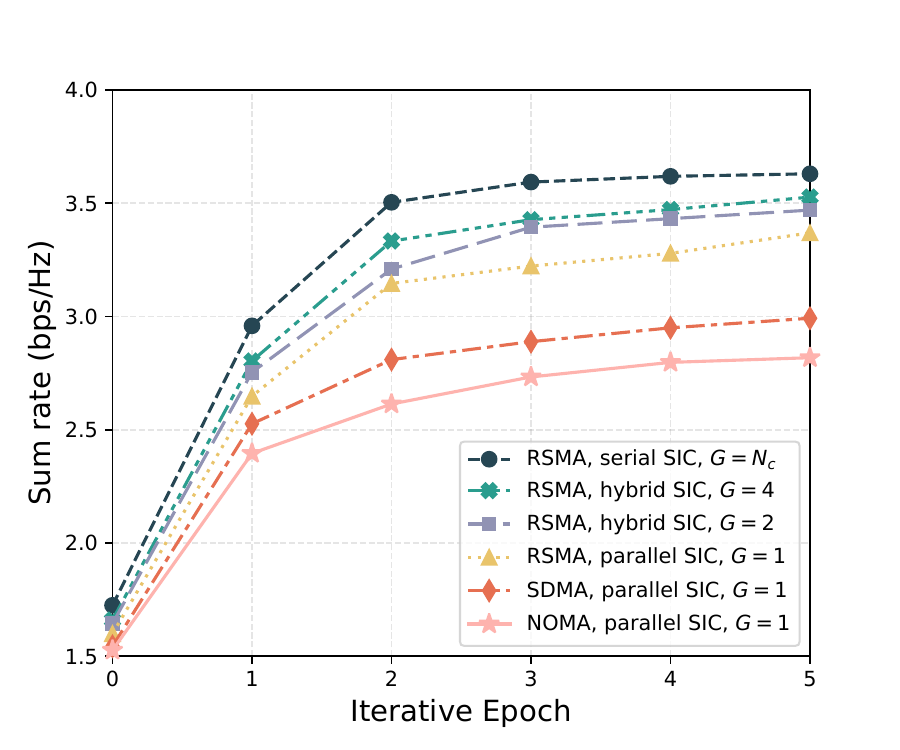}
\caption{Sum rate comparisons of RSMA, SDMA and NOMA with the corresponding SIC at SNR = 20 dB versus iterative epoch.}
\label{fig_3}
\end{figure}
Fig. 2 compares the performance of RSMA, SDMA, and NOMA with different SIC structures\footnote{To the hybrid SIC, we consider that the subcarriers are evenly split into each cluster for simplicity.}. RSMA demonstrates consistent superiority over SDMA and NOMA across all SNRs with parallel SIC. Though the serial SIC structure assists RSMA to achieve a higher sum rate, it brings too much complexity due to $G=N_c$. Indeed, we only set $G=2,4$ for the hybrid SIC structure and improve the system capacity compared with the parallel SIC. When $G=4$, RSMA achieves a sum rate of 4.15 bps/Hz at SNR = 25 dB, which is very close to the optimal serial SIC of 4.25 bps/Hz. 

Fig. 3 illustrates the sum rate trend during the alternative optimization at SNR = 20 dB. With our proposed hybrid precoding strategy, the system can converge within only about 5 iterative epochs. Similar results as Fig. 1 show the superiority of our proposed hybrid SIC structure.

Finally, we give the computation complexity of the hybrid SIC structure for OFDM-RSMA in Table I. The hybrid SIC architecture demonstrates remarkable efficiency with only 25\% ($G=2$) and 37.5\% ($G=4$) complexity increments relative to parallel SIC. This complexity-adaptive scheme enables flexible and dynamic reconfiguration across the serial-parallel design spectrum to meet different constraints.
\begin{table}[]
\caption{Complexity of Different SIC Structures for OFDM-RSMA\label{table1}}
\renewcommand\arraystretch{1.2}
\centering
\setlength{\tabcolsep}{3mm}{
\small
\begin{tabular}{c|c}
\hline
\textbf{Structure}             & \textbf{Computation Complexity}      \\ \hline
Hybrid     & $O(\frac{3G-1}{G}KN_tN_c^2+\frac{3G-1}{2G}KN_c^2)$       \\
Parallel, $G=1$   & $O(2KN_tN_c^2+KN_c^2)$      \\
Hybrid, $G=2$ & $O(\frac{5}{2}KN_tN_c^2+\frac{5}{4}KN_c^2)$        \\
Hybrid, $G=4$ & $O(\frac{11}{4}KN_tN_c^2+\frac{11}{8}KN_c^2)$ \\
Serial, $G=N_c$ & $O(3KN_tN_c^2+\frac{3}{2}KN_c^2)$
\\ \hline
\end{tabular}}
\end{table}

\section{Conclusion}
In this paper, a novel RSMA-assisted transceiver-coordinated scheme for ICI mitigation in MIMO-OFDM systems was proposed. We introduced a complexity-adaptive hybrid SIC receiver, where subcarriers were clustered to enable parallel intra-cluster and serial inter-cluster processing, achieving effective interference cancellation. A matching hybrid precoding strategy was developed at the transmitter, jointly optimizing analog/digital precoders through iterative methods. The proposed framework enhanced the system sum rate significantly with marginal complexity overhead.



%

\bibliographystyle{IEEEtran}
\bibliography{IEEEabrv,arxiv}

\end{document}